\documentstyle[12pt]{article}
\title{\bf The Unique Properties of $\omega$ Centauri Seen Through Str\"omgren
Eyes\footnote{Based on data collected at the European Southern Observatory (La
Silla, Chile)} }
\author{M.~Hilker $^1$ and T.~Richtler$^2$\\
\vspace{1cm}\\
\normalsize $^1$Departamento de Astronom\'\i a y Astrof\'\i sica, 
P.~Universidad Cat\'olica, Chile\\
\normalsize $^2$Departamento de Astronom\'\i a y Astrof\'\i sica,
Universidad de Concepcion, Chile}
\date{\mbox{}}
\begin{document}
\input{psfig.sty}
\maketitle
\pagestyle{empty}
%
%
\def\bull{\vrule height .9ex width .8ex depth -.1ex}
\makeatletter
\def\ps@plain{\let\@mkboth\gobbletwo
\def\@oddhead{}\def\@oddfoot{\hfil\tiny\bull\quad
``The Galactic Halo: from Globular Clusters to Field Stars'';
35$^{\mbox{\rm rd}}$ Li\`ege\ Int.\ Astroph.\ Coll., 1999\quad\bull}%
\def\@evenhead{}\let\@evenfoot\@oddfoot}
\makeatother
%
%
\def\beginrefer{\section*{References}%
\begin{quotation}\mbox{}\par}
\def\refer#1\par{{\setlength{\parindent}{-\leftmargin}\indent#1\par}}
\def\endrefer{\end{quotation}}
\def\farcm{\hbox{$.\mkern-4mu^\prime$}}
\def\farcs{\hbox{$.\!\!^{\prime\prime}$}}
%
%
{\noindent\small{\bf Abstract:} 
A revised metallicity calibration of the Str\"omgren $(b-y),m_1$ diagram,
based on a sample of globular cluster and field red giant stars, is presented.
This new calibration has been used to determine Str\"omgren metallicities
([Fe/H]$_{\rm phot}$) for more than 1400 red giants in $\omega$ Centauri.
Most of the stars belong to a 
metal poor ([Fe/H]$_{\rm phot} = -$1.7 dex), old population.
A second more metal rich ([Fe/H]$_{\rm phot} = -$0.9 dex) population
is found to be about 3 Gyr younger and more centrally concentrated than the 
old one. The CN-rich stars do not seem to follow the radial distribution of the
metal rich population.
}
%
%
\section{Introduction}

The globular cluster $\omega$ Centauri is known as an extraordinary 
object among the clusters
of our Galaxy. It is not only the most massive and flattened Galactic 
globular cluster, but also shows strong variations in the abundances of 
CNO elements as well as in iron.

Whereas the CNO variations can be explained by evolutionary mixing effects
in the stellar atmosphere as well as by primordial mixing, the iron abundance
variations need another explication (e.g. Vanture et al. 1994, Norris \& 
Da Costa 1995). Calcium abundance measurements of more than 500 red 
giants (Norris et al. 1996) show a bimodal metallicity distribution.
About 80\% of the stars have a low metallicity, the rest a 0.5 dex higher one.
The interpretation is two succesive epochs of star formation after the self 
enrichment within the cluster over a relatively lengthy period.
However, dynamical analysis of 400 stars in this sample (Norris et al. 
1997) shows rotation in the metal poor component, whereas the metal
rich one is not rotating. This favours a merger of two globular clusters with 
different masses, as shown by the model calulations of Makino et al. 
(1991).

Clearly, further analyses with large homogeneous samples are needed to
understand the complex formation and enrichment history of this outstanding
object.

\subsection{Why Str\"omgren photometry?}

Str\"omgren photometry has been proven to be a very useful metallicity 
indicator for globular cluster giants and subgiants (Richtler 1989,
Grebel \& Richtler 1992). The location of late type stars in the
Str\"omgren $(b-y),m_1$ diagram is correlated with their metallicities,
especially with their iron and CN abundances.

In the case of $\omega$ Centauri, CCD Str\"omgren photometry offers the
possibility to determine metallicities and study iron and CNO abundances
of a large number of stars simultaneously.

\section{Observations and reduction}

The observations have been performed in two observing runs in the nights 
11-15 May 1993 and 21-24 April 1995
with the Danish 1.54m telescope at ESO/La Silla. The CCD in use
was a Tektronix chip with 1024$\times$1024 pixels.
The $f$/8.5 beam of the telescope provides a scale of $15\farcs7$/mm,
and with a pixel size of 24 $\mu$m the total field is $6\farcm3 \times 
6\farcm3$. In total, 33 different fields have been observed in $\omega$ Cen
(18 with short exposures, 15 with long exposures), 4 fields in M22 and 5 fields
in M55. Figure~1 illustrates the selected fields in $\omega$ Cen.

The CCD frames were processed with the standard IRAF routines, instrumental
magnitudes were derived using DAOPHOT II.
After the photometric reduction, the matching of all frames, and calibration
of the magnitudes, the average photometric errors for the red giants used 
for the metallicity determination are 0.011 mag for $V$, 0.016 mag for $(b-y)$
and 0.023 mag for $m_1$. More details are presented in Richter et al. (1999).

\section{Revised metallicity calibration}

Red giants in the globular clusters $\omega$
Cen, M55, and M22 together with field giants
from Anthony-Twarog \& Twarog (1998) have been used to revise the metallicity
calibration of the Str\"omgren $(b-y),m_1$ diagram by Grebel \& Richtler
(1992). For all giants, accurate and homogeneous iron abundances
from high resolution spectroscopy are available in the literature.
In total, 58 CN-weak giants have been used. CN-rich stars have been excluded,
since their $m_1$ value mimics a too high iron abundance in the $(b-y),m_1$
diagram.
In order to cover a wide metallicity range, $-2.0 <$[Fe/H]$< 0.0$ dex, our
new calibration is connected to a previous calibration by
Grebel \& Richtler (1992) around solar metallicities.
In the color range  $0.5 < (b-y)
< 1.1$ mag, for which our calibration is valid, the loci of equal iron
abundances lie on straight lines.

\begin{figure}[t]
\begin{minipage}{8cm}
\psfig{figure=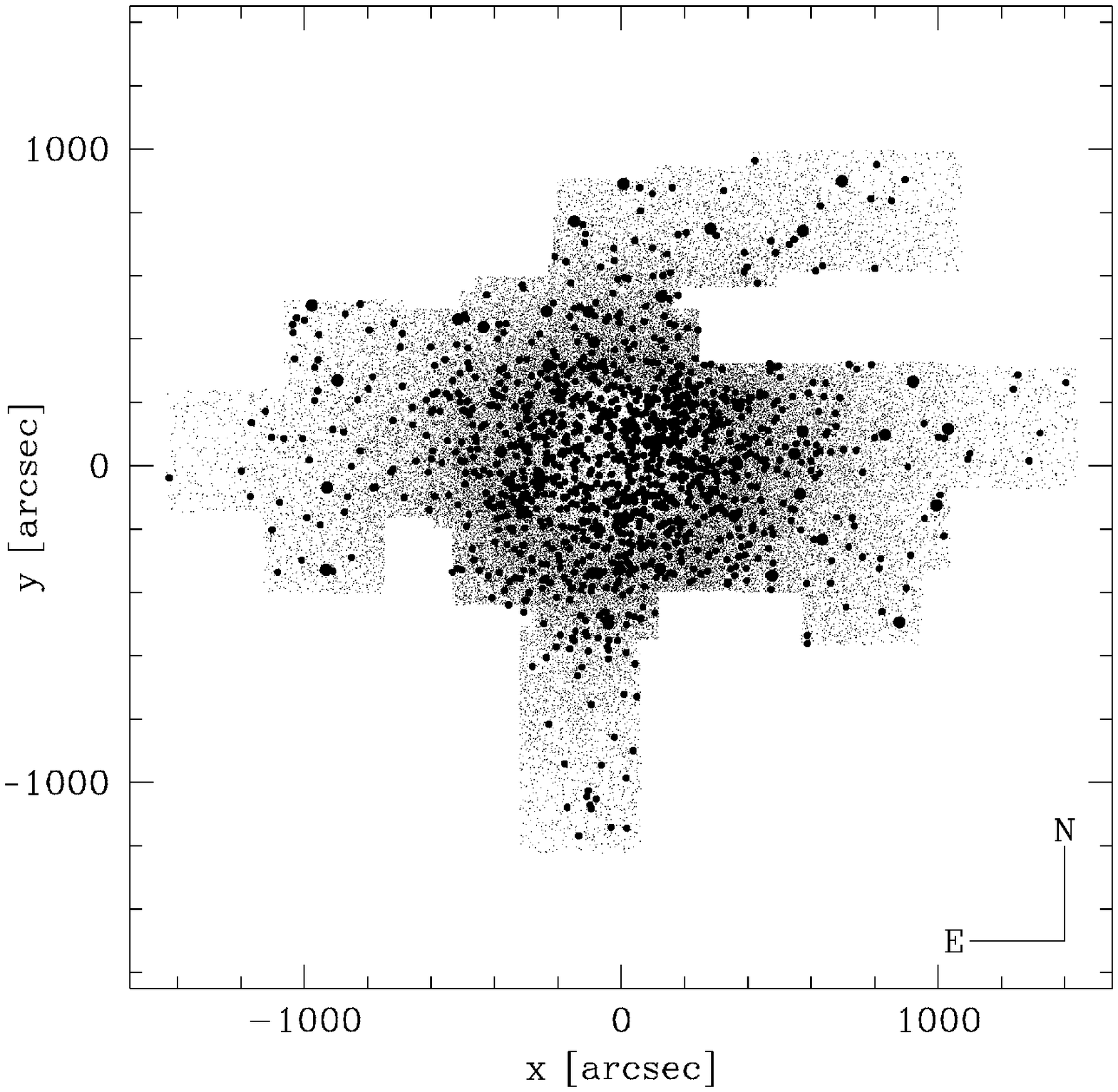,height=8.0cm,width=8.0cm
,bbllx=18mm,bblly=65mm,bburx=190mm,bbury=231mm}
\vspace{0.3 cm}
\end{minipage}
\hfill
\begin{minipage}{8cm}
\psfig{figure=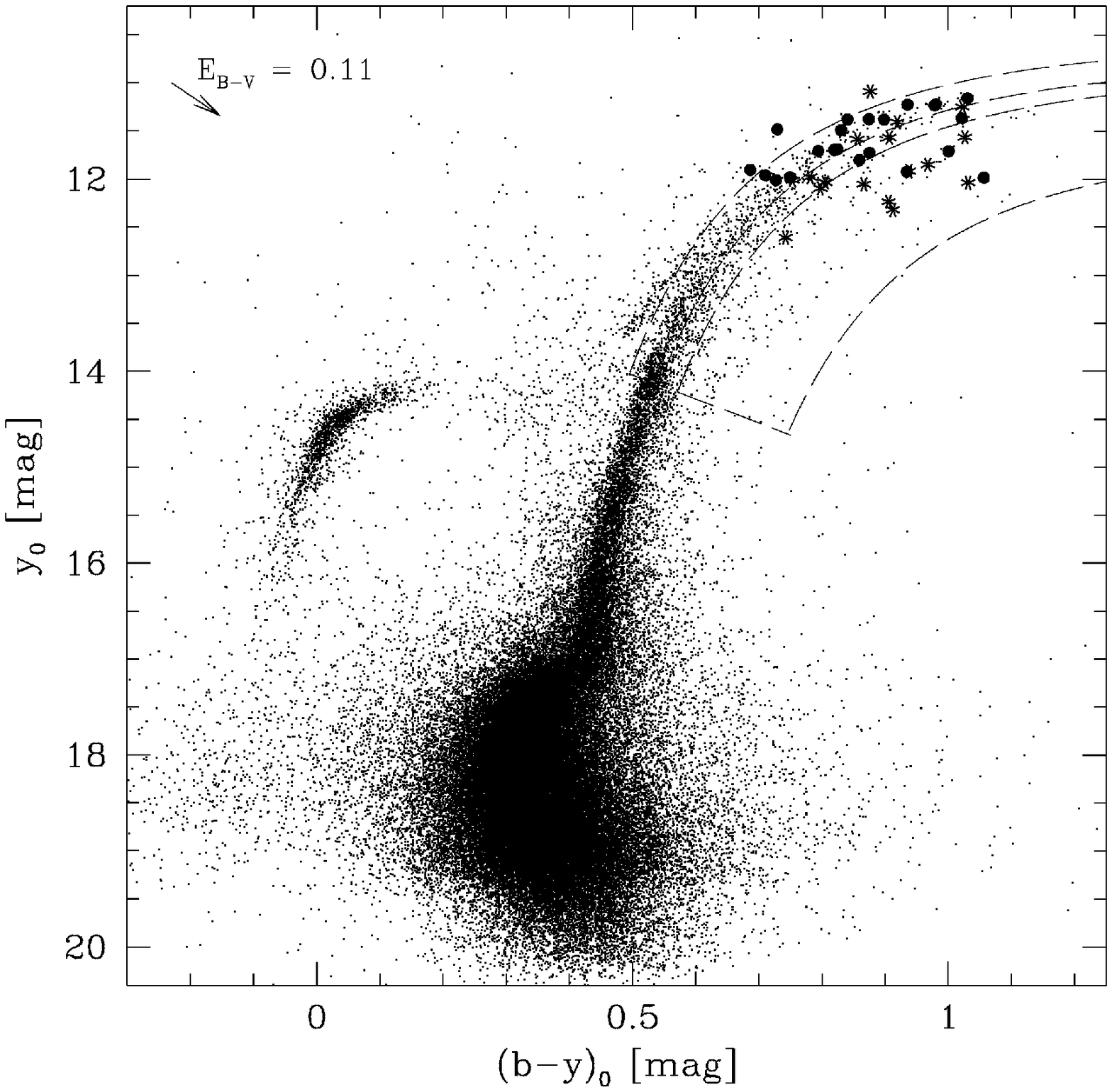,height=8.0cm,width=8.0cm
,bbllx=18mm,bblly=65mm,bburx=190mm,bbury=231mm}
\vspace{0.3 cm}
\end{minipage}
\begin{minipage}{8cm}
\caption{Position plot of all observed stars with a $V$ magnitude brighter 
than 19.0 mag and a photometric error less than 0.1 mag. Bold dots indicate 
the position of selected red giants (see Fig.~2), large bold dots the ones 
with known spectroscopic abundances used in the metallicity calibration.}
\end{minipage}
\hfill
\begin{minipage}{8cm}
\caption{CMD of all stars with $\sigma_{rm phot} < 0.1$ mag. Filled circles 
indicate red giants with normal CN abundances, whereas stars marked with 
asterisks are known to be CN-strong. All stars within the area enclosed by 
dashed lines have been used for further analysis.}
\end{minipage}
\end{figure}

Following the calibration by Grebel \& Richtler, a relation of the form\\
\centerline {[Fe/H] $= (m_1 + a_1 \cdot (b-y) + a_2)/(a_3 \cdot (b-y) +
a_4)$}\\
has been chosen for the fit. The derived coefficients are

$a_1 = -1.277\pm0.050$ \qquad \qquad $a_2 = \;\;\; 0.331\pm0.035$

$a_3 = \;\;\; 0.324\pm0.035$ \qquad \qquad $a_4 = -0.032\pm0.025$\\

The dispersion of the difference between the Str\"omgren metallicity and
spectroscopic iron abundances is 0.11 dex, which indicates the average accuracy
of the new calibration for a single giant. The detailed analysis of the 
metallicity calibration will be published in Hilker (1999).

\begin{figure}[t]
\begin{minipage}{8cm}
\psfig{figure=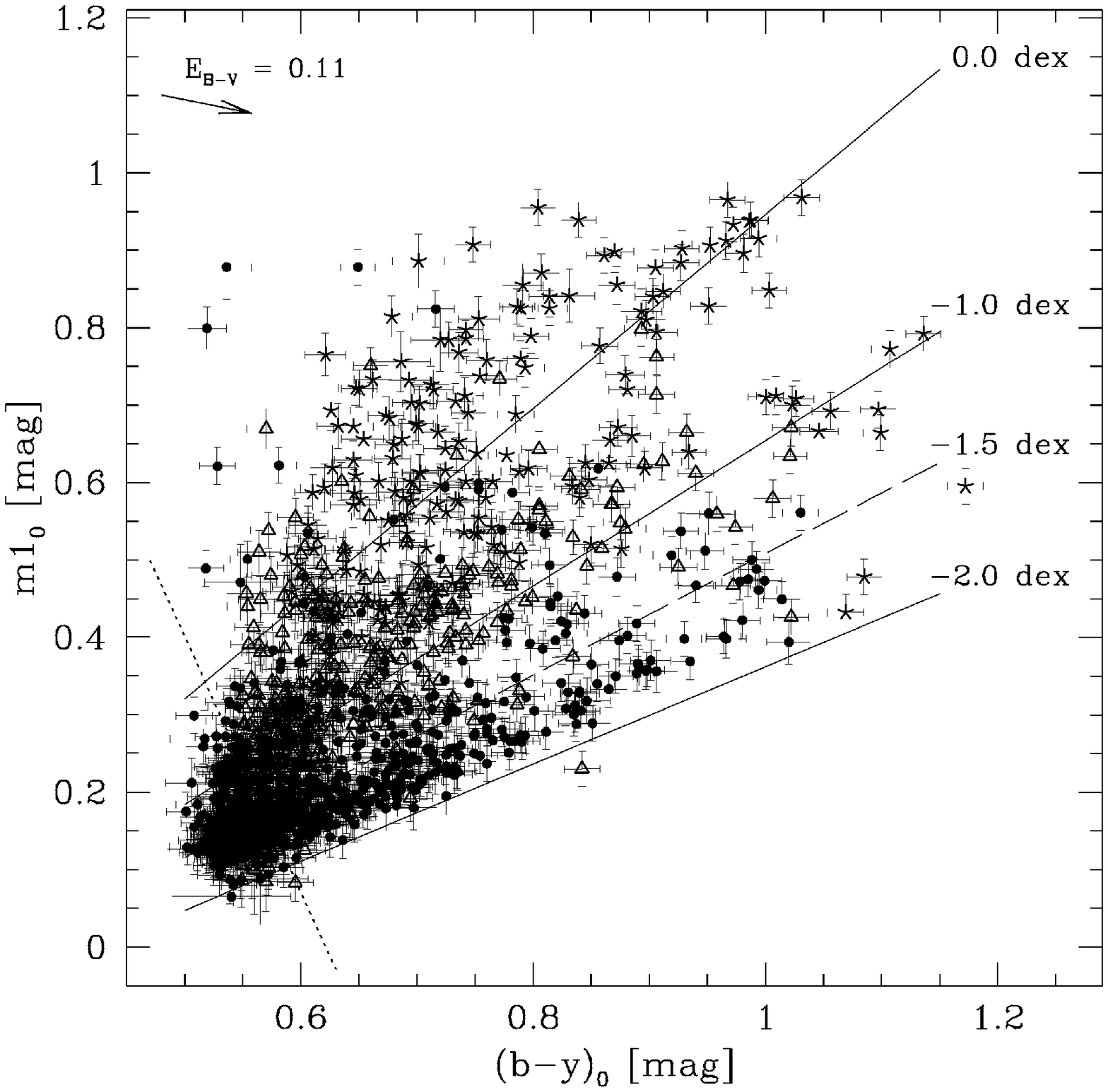,height=8.0cm,width=8.0cm
,bbllx=18mm,bblly=65mm,bburx=190mm,bbury=231mm}
\vspace{0.3 cm}
\end{minipage}
\hfill
\begin{minipage}{8cm}
\psfig{figure=mh_fig4.ps,height=8.0cm,width=8.0cm
,bbllx=9mm,bblly=65mm,bburx=195mm,bbury=246mm}
\vspace{0.3 cm}
\end{minipage}
\begin{minipage}{8cm}
\caption{The $(b-y),m_1$ diagram for more than 1400 selected giants (see 
Fig.~2) is shown together with the lines of constant metallicity from the
new calibration. The error bars include photometric and calibration errors.
Dots indicate stars from the blue side of the RGB, triangles the ones from 
the red RGB side, and asterisks all stars lying apart from the ``main RGB''
(see Fig.~2).}
\end{minipage}
\hfill
\begin{minipage}{8cm}
\caption{In this plot the metallicity distribution of more than 1000 red giants
and subsamples is shown. Only stars that are redder than the dotted line in 
Fig.~3 have been selected. The hashed
histograms are the distributions of the ``blue'' RGB stars (metal poor), ``red''
RGB stars (more metal rich), and stars redder than the ``main RGB'', most of
them being CN-rich.}
\end{minipage}
\end{figure}

\section{Analysis and Results}

Figure 2 shows the color magnitude diagram of all observed stars
in $\omega$ Cen with a photometric error less than 0.1 mag ($\simeq$ 92500 
stars). The colors have been corrected for a reddening value of $E_{B-V} =
0.11$ mag (Zinn 1985). Bold dots and asterisks indicate the stars with known 
spectroscopic abundances which have been used for the metallicity calibration.
The broad red giant branch at magnitudes brighter than $V = 14$ mag cannot be
explained by photometric errors, but is due to a spread in age and
metallicity. 
All stars in the CMD which are enclosed by the dashed lines were used
for the metallicity determination ($\simeq 1400$ stars). For further analysis
the RGB was divided into three subsamples: a blue and a red side of the 
RGB, and all stars that are redder than the ``main'' RGB. 

\begin{figure}[t]
\hfill
\begin{minipage}{8cm}
\psfig{figure=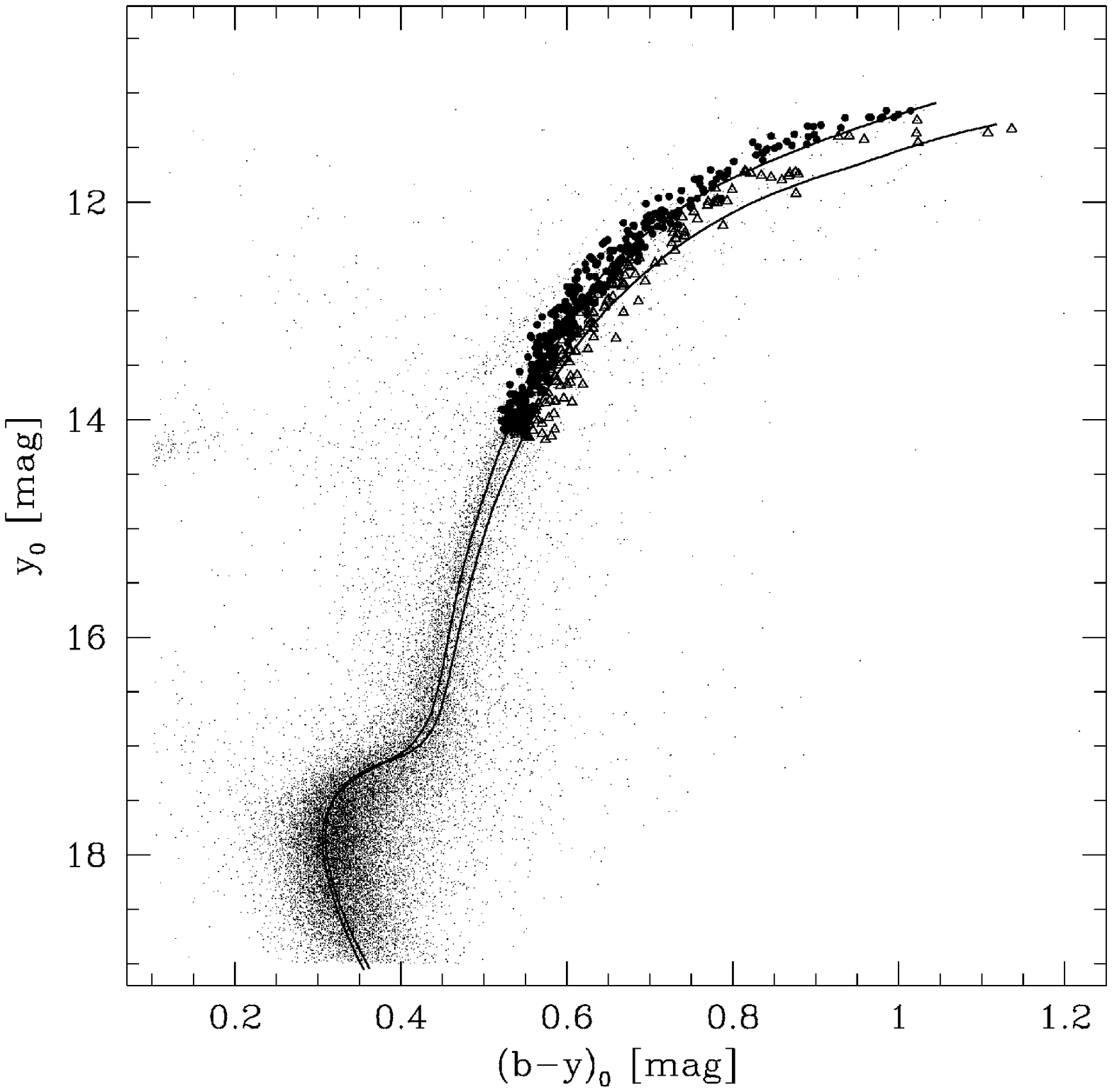,height=8.0cm,width=8.0cm
,bbllx=25mm,bblly=65mm,bburx=195mm,bbury=231mm}
\vspace{0.3 cm}
\end{minipage}
\hfill
\begin{minipage}{8cm}
\psfig{figure=mh_fig6.ps,height=8.0cm,width=8.0cm
,bbllx=9mm,bblly=65mm,bburx=195mm,bbury=246mm}
\vspace{0.3 cm}
\end{minipage}
\begin{minipage}{8cm}
\caption{In this CMD isochrones from Bergbusch \& VandenBerg (1992), converted
to Str\"omgren colors by Grebel \& Roberts (1994), have been fitted to the
metal poor and more metal rich subpopulations in $\omega$ Cen. The
corresponding ages and metallicities are indicated in the panel.}
\end{minipage}
\hfill
\begin{minipage}{8cm}
\caption{This plot shows the cumulative distribution of different subsamples of
red giants in a $6\farcm7$ wide strip west of the center of $\omega$ Cen.
The selections are indicated in the panel. Clearly the metal richer stars
are more concentrated than the old metal poor population.}
\end{minipage}
\end{figure}

The metallicity spread is illustrated in the $(b-y),m_1$ diagram (Fig.~3) and
the metallicity histogram (Fig.~4). 
Most of the blue RGB stars populate a prominent iso-metallicity line around
$-1.7$ dex. These stars are not affected by CN enrichment and form a single
metal poor population. Stars from the red side of the RGB have 
metallicities mainly in the range $-1.3$ to $-0.5$ dex. They appear well
separated from the metal poor population. Stars with Str\"omgren
metallicities higher than about $-0.8$ dex are supposed to be CN-rich stars 
of one of the two populations. However, lots of them are redder than the 
``main'' RGB. Since their $(b-y)$ color is not influenced by CN variations,
their existence in the CMD can only be explained by higher iron abundances.
Therefore we conclude that there exists a true tail towards higher 
metallicities not only in the CN but also in the iron abundances.

Note that the numbers in the metallicity histogram do not represent the right 
proportion of metal poor to metal rich stars, since the stars have been 
selected due to a high accuracy in metallicity (all stars redder than the 
dotted line in Fig.~3) and not due to a mass cut in the CMD. When accounting 
for a mass selected sample the ratio of metal poor to metal rich stars is
about 3:1.

The different metallicities of the two populations have been used to 
estimate their ages. Since newly calculated isochrones in the Str\"omgren
system still are missing, isochrones from Bergbusch \& VandenBerg (1992),
converted to Str\"omgren colors by Grebel \& Roberts (1994), have been
used. These isochrones were found to represent the shape of the RGB best,
and therefore are very useful to determine relative ages.
In Fig.~5 the two best fitting isochrones are shown. The metal poor stars
($-1.85<$[Fe/H]$<-1.55$, bold dots) are best representated by an isochrone
of 18 Gyr and $-1.68$ dex, the metal richer ones ($-1.2<$[Fe/H]$<-0.8$,
triangles) by an isochrone of 15 Gyr and $-1.28$ dex. The uncertainty in the
age determination is about 1 Gyr. An isochrone of 18 Gyr for the red RGB stars 
cannot reproduce the narrow appearance of the CMD in the fainter RGB and 
turnover region. The fitted iron abundance is somewhat lower than the 
Str\"omgren metallicity of the metal rich population. This is due to the fact
that CN-rich stars already influence the measurements in this metallicity
range and pretend a too high metallicity.

The radial distribution of the different subpopulations in $\omega$ Cen
is shown in Fig.~6. Only stars within a $6\farcm7$ wide strip west of
the center of $\omega$ Cen have been selected, since these fields 
belong to the most homogeneous set of long exposures. The East-West axis is 
perpendicular to the rotation axis of the cluster. The metal richer and younger 
stars from the red side of the RGB are clearly more concentrated than the old 
metal poor population within 10 arcmin radius from the cluster center.
The situation is not that clear for the CN-rich stars. They seem to follow
more the distribuion of the metal poor stars than that of the metal rich ones.
However, this result has to be taken with caution, since number statistics
are low, and this sample of stars most probably is a mix of CN enriched stars 
from both populations, and even foreground stars.

Putting all the facts from the metallicity determination, age estimation,
and spatial distribution analyses together, there seems to be no doubt that 
$\omega$ Cen consists of (at least) two distinct populations of stars, which
gives this extraordinary object more the characteristics of a galaxy than a
globular cluster.

\section{Summary}

A new calibration of the Str\"omgren $(b-y),m_1$ diagram to the iron abundance
of red giants has been presented. The Str\"omgren metallicity calibration is 
valid in the abundance range $-2.3 <$ [Fe/H] $< 0.0$ dex.

For Str\"omgren metallicities higher than $-$1.0 dex, CN-weak stars cannot be 
distinguished in the $(b-y),m_1$ diagram from stars with lower iron abundances
but higher CN bands strength.

For more than 1400 red giants Str\"omgren metallicities have been determined.
Besides a main metal poor ([Fe/H]$_{rm phot} = -$1.7 dex) population, a more 
metal rich population ([Fe/H]$_{rm phot} = -$0.9 dex), and a extended tail 
of CN-rich stars with higher Str\"omgren metallicities have been found.

The metal poor population is old. The metal richer population is about
3 Gyr younger and more centrally concentrated than the old one.

Our findings are consistent with a scenario in which self enrichment
within the cluster has been taken place over a period of about 2-4 Gyr
with subsequent new star formation from the enriched and more concentrated
material.

We point out that the conditions for such an enrichment history can be
found in nuclei of dwarf galaxies, where the gas can be easily retained and a
secondary star formation is common. The capture and dissolution of a
nucleated dwarf galaxy by our Milky Way and the survival of $\omega$ Cen
as its nuclues would be an attractive explanation for this extraordinary
object.

%
%
\section*{Acknowledgements}
We thank Boris Dirsch and Philipp Richter for helpful discussions and
useful comments.
This research was supported through `Proyecto FONDECYT 3980032'.
%
%
 
\beginrefer

\refer  Anthony-Twarog B.J., Twarog B.A., 1998, AJ 116, 1922

\refer Bergbusch P.A., VandenBerg D.A., 1992, ApJS 81, 163

\refer Grebel E.K., Richtler T., 1992, A\&A 253, 359

\refer Grebel E.K., Roberts W.J., 1995, AAS 186, 0308

\refer Hilker M., 1999, A\&A submitted

\refer Makino J., Akiyama K., Sugimoto D., 1991, ApSS 185, 63

\refer Norris J.E., Da Costa G.S., 1995, ApJ 447, 680

\refer Norris J.E., Freeman K.C., Mayor M., Seitzer P., 1997, ApJ 487, L187

\refer Norris J.E., Freeman K.C., Mighell K.J., 1996, ApJ 462, 241

\refer Richter P., Hilker M., Richtler T., 1999, A\&A in press

\refer Richtler T., 1989, A\&A 211, 199

\refer Twarog B.J., Twarog B.A., 1998, AJ 116, 1922

\refer Vanture A.D., Wallerstein G., Brown J.A., 1994, PASP 106, 835

\refer Zinn R., 1985, ApJ 293, 424

\endrefer           
\end{document}